# La$_3$Pd$_2$NaO$_9$: A High-Valent Insulating Palladate

Qingqing Yang,[1] Ning Guo,[2] Tieyan Chang,[3] Zheng Guo,[1] Xiaoli Wang,[1] Chuanyan Fan,[1] Chao Liu,[1] Lu Han,[1] Feiyu Li,[1] Tao He,[1] Qiang Zheng,[2] Yu-Sheng Chen,[3] Junjie Zhang[1]*

[1]Institute of Crystal Materials, State Key Laboratory of Crystal Materials, Shandong University, Jinan 250100, Shandong, China

[2]CAS Key Laboratory of Standardization and Measurement for Nanotechnology, CAS Center for Excellence in Nanoscience, National Center for Nanoscience and Technology, Beijing 100190, China

[3]NSF's ChemMatCARS, The University of Chicago, Argonne, IL 60439, USA

*Email: junjie@sdu.edu.cn

**Abstract**: A high-valent palladate, La$_3$Pd$_2$NaO$_9$, has been synthesized for the first time. Single crystals with dimensions of 20 μm on edge were successfully grown using the flux method at 420 °C and 70 bar oxygen pressure. Energy dispersive spectroscopy (EDS) and inductively coupled plasma mass spectroscopy (ICP) measurements show that the atomic ratio of La: (Pd+Na) is 3: 3 and Pd: Na is 2: 1. X-ray photoelectron spectroscopy (XPS) measurements show that the oxidation state of Pd is dominated by +4. Synchrotron X-ray single-crystal diffraction measurements revealed that this material crystallizes in the monoclinic $P2_1/c$ space group with charge ordering of Na and Pd. Real-space imaging via scanning transmission electron microscopy (STEM) confirmed the crystal structure and revealed excellent sample homogeneity. Electrical resistivity measurements show an insulating behavior. Magnetic measurements show an unexpected paramagnetic behavior, which probably originate from a small fraction of high-spin Pd$^{2+}$ evidenced by XPS. The successful growth of La$_3$Pd$_2$NaO$_9$ single crystals with a high-valent oxidation state of Pd offers an approach for exploring interesting palladates, including potential bilayer Ruddlesden-Popper palladates analogous to the high temperature superconducting La$_3$Ni$_2$O$_7$.

Keywords: high-valent palladate; charge ordering; high pO$_2$ crystal growth; flux growth; perovskite.

## 1. Introduction

Since the discovery of cuprate high-temperature superconductors in 1986,[1] clarifying the mechanism of high-temperature superconductivity has been a challenge in the fields of condensed matter physics and materials science.[2-5] One strategy is to look for cuprate analogs to explore more high-temperature superconducting systems.[6] Ni, which is next to Cu in the periodic table, has attracted a lot of attention.[7-13] The report of superconductivity in Nd$_{0.8}$Sr$_{0.2}$NiO$_2$ thin films in 2019 triggered a research boom in nickelate superconductivity,[14] and since then more nickel-based thin film superconductors have been reported.[15-19] These films are obtained by topotactic reduction of corresponding Ruddlesden-Popper (R-P) phases R$_{n+1}$Ni$_n$O$_{3n+1}$ (R being a lanthanide). Among them, the material with the highest superconducting transition temperature ($T_c$) is (Sm-Eu-Ca-Sr)NiO$_2$ with ~ 37 K.[19] In 2023, bulk single crystals of the bilayer R-P phase La$_3$Ni$_2$O$_7$ were reported to show signature of superconductivity with a transition temperature close to 80 K at pressures higher than 14 GPa,[20-22] making it a potential material for high-temperature superconductivity. Later, the trilayer R$_4$Ni$_3$O$_{10}$ was also reported to superconduct under high pressures.[23-27] In addition, bulk nickelates with new crystal structures were designed and synthesized, including hybrid R-P nickelates La$_2$NiO$_4$·La$_3$Ni$_2$O$_7$[28] and La$_2$NiO$_4$·La$_4$Ni$_3$O$_{10}$,[29-31] the former of which has been observed to have a $T_c$ ~ 64 K under high pressure.[32] Recently, La$_3$Ni$_2$O$_7$ thin films annealed in ozone were also reported to show signature of superconductivity with a $T_c$ ~ 45 K at ambient pressure.[33, 34] More recently, by introducing chemical pressure, Li et al. succeeded in growing high quality single crystals of La$_2$SmNi$_2$O$_7$ at ambient pressure, and pushing the superconducting transition temperature to above 90 K, which is the highest among the nickelate superconductors.[35]

A natural question is whether there exist superconductors in palladates since Pd is just locating below Ni in the next periodic row. Palladates have been predicted to be superconductors for a long time.[36-41] The oxidation state of palladium in most predicted palladate superconductors is monovalent. Based on theoretical calculations, the $Pd^+$-4d orbitals reside between the 3d orbitals of $Cu^{2+}$ and $Ni^+$ in the energy level diagrams, and closer to $Cu^{2+}$ compared to $Ni^+$.[40] The larger core charge of Pd strongly attracts electrons, and the larger band-width in $Pd^+$-4d lead to stronger hybridization with the 2p orbitals of $O^{2-}$ than $Ni^+$-O.[40] Transition metal and oxygen hybridization is important for high-$T_c$ superconductors,[6, 42] thus palladates are promising candidates to explore. In real materials, $Pd^+$-oxides are very rare. Only $PdAO_2$ (A = Co, Cr, Rh),[43-45] whose structure is a stacked arrangement of $Pd^+$ metallic triangular layers and $AO_2$ layers, have been reported. Up to date, no palladates with square lattice of $Pd^+$ has been synthesized.

A feasible way to prepare $Pd^+$-oxides is topotactic reduction of R-P palladates $R_{n+1}Pd_nO_{3n+1}$ (R being a lanthanide), similar to nickelates[46] and iron oxides.[47] Along this line, square-lattice $LaPdO_2$ and $La_4Pd_3O_8$ has been proposed via reducing $LaPdO_3$ and $La_4Pd_3O_{10}$, respectively.[36] For R-P palladates, only n = 1 member $(R_2PdO_4)$[48] and n = ∞ member $(LaPdO_3)$[49] have been synthesized. $R_2PdO_4$ with T'-structure, the same as $Nd_2CuO_4$,[50] has been reported for a variety of lanthanide ions and both polycrystalline samples and thin films (unfortunately no single crystals have been reported up to date) have been synthesized.[48, 51, 52] These single-layer palladates are insulators and exhibit diamagnetic behavior, consistent with the low spin $d^8$ configuration.[48, 51, 52] The resistivity decreases after electron doping by substituting $La^{3+}$ with $Ce^{4+}$, but it still shows semiconducting behavior with temperature.[51, 52] One exception is the thin film samples annealed in vacuum, where a metal-insulator transition was observed.[52] For n = ∞, polycrystalline $LaPdO_3$ with paramagnetic behavior was synthesized under a high pressure of 5 GPa.[49] The electronic configuration of Pd has been proposed to be $t_{2g}^6\sigma^{*1}$, which should exhibit metallic behavior. Later, Penny G[53] measured the resistivity of high-pressure synthesized $LaPdO_3$ pellet containing KCl and impurities such as $La_2Pd_2O_5$ and PdO, and they claimed that $LaPdO_3$ is metallic due to its low resistivity (< 0.1 Ω cm below 350 K) although a small increase of resistivity with decreasing temperature is seen. No other R-P palladates have been reported up to date due to the dominance of $La_4PdO_7$ and $La_2Pd_2O_5$ in the thermodynamic phase diagram.[54]

In this contribution, we attempted to grow $LaPdO_3$ crystals using NaOH-KOH flux at $pO_2$ = 70 bar and T = 420 °C for subsequent topotactic reduction to obtain $LaPdO_2$. Unexpectedly, 1/3 of the Pd site was substituted by Na, resulting in a high valence palladate, $La_3Pd_2NaO_9$, evidenced by EDS, ICP and XPS data. XPS measurement shows that the oxidation state of Pd is dominated by +4 with a small amount of +2. Real-space imaging via STEM revealed a perovskite-like structure and excellent sample homogeneity. Synchrotron X-ray single-crystal diffraction determined that the material crystallizes in the monoclinic $P2_1/c$ space group with charge ordering of Na and Pd. Resistivity measurements show that $La_3Pd_2NaO_9$ is an insulator, which is consistent with charge ordering of Na and Pd. The magnetization data show an anomalous paramagnetic behavior, which can be attributed to the presence of a small fraction of high-spin $Pd^{2+}$ as evidenced by XPS. The successful preparation of $La_3Pd_2NaO_9$ opens a door for preparing single crystals of other interesting palladates, including potential bilayer R-P palladates analogous to the high temperature superconducting $La_3Ni_2O_7$.

## 2. Experimental Section
**2.1 High $pO_2$ Flux Growth of $La_3Pd_2NaO_9$ Crystals.** All experiments were carried out at a high-pressure flux furnace designed for operation at $T_{max}$ = 500 °C and $P_{max}$ = 160 bar (Xi'an Taikang Biotechnology Co., Ltd). $La_2O_3$ (Sigma Aldrich, 99.99%), Pd (Aladdin, 99.5%), NaOH (Aladdin, 99.9%) and KOH (Aladdin, 99.99%) were used as raw materials. $La_2O_3$ was baked at 650 °C for 10 h before use. Powders of $La_2O_3$ and Pd with a stoichiometric ratio were weighed, mixed, and ground and then it was mixed with NaOH-KOH (molar ratio 1:1) flux in a glove box at a weight ratio of 1:5, 1:10, or 1:20 (see **Table S1**). The mixture was placed in an $Al_2O_3$ crucible (Φ30 mm × 30 mm), which was then covered with a flat lid and put into the high-pressure furnace. The furnace was purged with high-purity oxygen several times. Afterwards, certain amount of high-purity oxygen was filled into the furnace at room temperature, so that when the furnace temperature reached 420 °C, the target $pO_2$ (e.g., 10, 30, 50, or 70 bar) was achieved inside the furnace. The furnace was maintained at targeted temperature and $pO_2$ for 48 h in order to dissolve raw materials. During this period, oxygen is consumed due to the reaction of raw materials to form $La_3Pd_2NaO_9$. Therefore, we connected the gas inlet to the oxygen cylinder and adjusted it through a pressure regulator to maintain the pressure at the expected value. After dwelling, the furnace was cooled down to 300 °C at the rate of 3.5, 3, 2.5, or 2 °C/h, followed by furnace cooling to room temperature. Finally, the remaining flux was removed with distilled water and black crystals with shiny facets were obtained.

**2.2 Powder X-ray Diffraction (PXRD).** A Bruker AXS D2 Phaser X-ray powder diffractometer was used to check phase purity. Data were collected at room temperature using Cu $K_\alpha$ radiation ($\lambda$ = 1.5418 Å) in the $2\theta$ range of 10-100° with a scan step size of 0.02° and a scan time of 0.5 s per step. TOPAS 6 was used for Rietveld refinement. Refinement parameters include the background (Chebychev function, order 11), zero shift, lattice parameters, crystallite size L, strain G, atomic positions, and thermal parameters.

**2.3 Scanning Electron Microscopy (SEM).** The morphology of the as-grown crystals was observed using a scanning electron microscope. The SEM images were obtained by a Zeiss Gemini SEM 500 microscope incident electron of 3.00 KV.

**2.4 EDS.** The Oxford UltimMax 170 energy spectrum on Zeiss Gemini SEM 500 was used for qualitative and quantitative analyses of the as-grown crystals.

**2.5 ICP.** The Inductively Coupled Plasma Mass Spectrometer PerkinElmer-Avio550 max was used for quantitative analysis of crystals. Crystals of 4.6 mg were thoroughly ground and dissolved in 5 mL of concentrated nitric acid (GR). The samples of 350, 500, 500, 800, and 800 μL were taken using a pipette gun and poured into a 100 mL volumetric flask. Distilled water was added to dilute and shake well. 10 mL of solution was taken from each volumetric flask as samples 1-5 for measurements. Blank was also prepared to ensure the accuracy of the test. 350, 500, 500, 800, and 800 μL nitric acid (GR) were taken using a pipette gun and poured into a 100 mL volumetric flask. Distilled water was added to dilute and shake well. 10 ml of solution was taken from each volumetric flask as blank samples 1-5 for measurements. The final concentration of each element tested was determined by subtracting the data of blank samples 1-5 from the data of samples 1-5.

**2.6 XPS.** The crystals were ground and pressed into a smooth, flat pellet for measurement. The oxidation state of Pd was measured using a PHI Genesis 500 X-ray photoelectron spectrometer with an Al $K_\alpha$ light source, a spot diameter of 200 μm, and a fine scan was performed in the range of 330-350 eV. The software Avantage was used to process the XPS measurement data.

**2.7 Synchrotron X-ray Single-Crystal Diffraction (SXRD).** SXRD data were collected using synchrotron radiation ($\lambda$ = 0.43060 Å) at GeoSoilEnviroCARS (Sector 13) at the Advanced Photon Source, Argonne National Laboratory. A single crystal with dimension of ~15 μm was mounted to the tip of glass fiber and measured using a Huber 3-circle diffractometer. Indexing, data reduction, and image processing were performed using Bruker APEX5 software.[55] The structure was solved by direct methods and refined with full matrix least-squares methods on $F^2$. All atoms were modeled using anisotropic ADPs, and the refinements converged for $I > 2\sigma(I)$, where $I$ is the intensity of reflections and $\sigma(I)$ is the standard deviation. Calculations were performed using the SHELXTL crystallographic software package.[55] Details of crystal parameters, data collection, and structure refinement at 296(2) K are summarized in **Table 1**, and atomic coordinateds and equivalent isotropic atomic displacement parameters are listed in **Table 2**. Further details of the crystal structure investigations may be obtained from the joint CCDC/FIZ Karlsruhe online deposition service by quoting the deposition number CSD 2445228.service by quoting the deposition number CSD 2445228.

Table 1. Crystal data and structure refinement for La$_3$Pd$_2$NaO$_9$.

| Empirical formula | La$_3$Pd$_2$NaO$_9$ |
| --- | --- |
| Formula weight | 796.73 |
| Temperature | 296(2) K |
| Crystal system | Monoclinic |
| Space group | P2$_1$/c |
| a | 16.777(5) Å |
| b | 5.8311(17) Å |
| c | 9.759(3) Å |
| α | 90° |
| β | 124.969(3)° |
| γ | 90° |
| Cell volume | 782.3(4) Å$^3$ |
| Z | 4 |
| Calculated density | 6.765 g/cm$^3$ |
| μ | 11.147 mm$^{-1}$ |
| F (000) | 1384.0 |
| Radiation | synchrotron X-ray (λ = 0.43106 Å) |
| 2θ range for data collection | 1.796 to 46.136° |
| Index ranges | -29 ≤ h ≤ 29, -9 ≤ k ≤ 9, -14 ≤ l ≤ 14 |
| Reflections collected | 10532 |
| Independent reflections | 3448 [R$_{int}$ = 0.0678, R$_{sigma}$ = 0.1802] |
| Data/restraints/parameters | 3448/0/144 |
| Goodness-of-fit on F$^2$ | 0.902 |
| Final R indexes [I ≥ 2σ (I)] | R$_1$ = 0.0575, wR$_2$ = 0.1847 |
| Final R indexes [all data] | R$_1$ = 0.0756, wR$_2$ = 0.2145 |
| Largest diff. peak/hole | 3.50/-3.27 e Å$^{-3}$ |

Table 2. Atomic coordinates occupancy and thermal vibration parameters of La$_3$Pd$_2$NaO$_9$.

| Site | Atom | x | y | z | Occ | U$_{eq}$ |
| --- | --- | --- | --- | --- | --- | --- |
| La1 | La | 0.92472(3) | 0.56136(7) | 0.25262(6) | 1 | 0.00718(18) |
| La2 | La | 0.74468(3) | 0.44213(7) | 0.74862(5) | 1 | 0.00726(18) |
| La3 | La | 0.41598(3) | 0.43997(7) | -0.24597(5) | 1 | 0.00760(18) |
| Pd1 | Pd | 0.67075(4) | 0.50300(7) | 0.00528(7) | 0.879(6) | 0.0049(2) |
| Pd2 | Pd | 0.82989(3) | 0.49798(7) | 0.49847(7) | 0.872(6) | 0.0048(2) |
| Pd3 | Pd | 1 | 0.5 | 1 | 0.265(7) | 0.0070(7) |
| Pd4 | Pd | 0.5 | 0.5 | -0.5 | 0.237(7) | 0.0059(7) |
| Na1 | Na | 0.67075(4) | 0.50300(7) | 0.00528(7) | 0.121(6) | 0.0049(2) |
| Na2 | Na | 0.82989(3) | 0.49798(7) | 0.49847(7) | 0.128(6) | 0.0048(2) |
| Na3 | Na | 1 | 0.5 | 1 | 0.735(7) | 0.0070(7) |
| Na4 | Na | 0.5 | 0.5 | -0.5 | 0.763(7) | 0.0059(7) |
| O1 | O | 0.7514(4) | 0.7985(8) | 0.4482(7) | 1 | 0.0103(10) |
| O2 | O | 0.7131(4) | 0.3001(8) | 0.4481(7) | 1 | 0.0106(9) |
| O3 | O | 0.5877(4) | 0.3098(9) | 0.0478(7) | 1 | 0.0140(10) |
| O4 | O | 0.9098(4) | 0.2116(9) | 0.5478(7) | 1 | 0.0138(10) |
| O5 | O | 0.9434(4) | 0.6901(9) | 0.5457(7) | 1 | 0.0134(10) |
| O6 | O | 0.6211(4) | 0.7834(9) | 0.0486(7) | 1 | 0.0128(10) |
| O7 | O | 0.5592(4) | 0.5259(9) | -0.2389(8) | 1 | 0.0150(11) |
| O8 | O | 0.7854(4) | 0.4703(8) | 0.2525(8) | 1 | 0.0117(10) |
| O9 | O | 0.8823(4) | 0.5258(9) | 0.7385(8) | 1 | 0.0118(10) |

**2.8 STEM.** The single crystals were crushed in ethanol, and drops of the suspensions were deposited on lacey carbon-coated molybdenum grids and dried in air for STEM observations. High-angle annular dark-field (HAADF)-STEM images were obtained at an accelerating voltage of 300 kV on an aberration-corrected transmission electron microscope (Spectra 300, Thermo Fisher Scientific), equipped with a field-emission electron gun. The probe convergence semi-angle and inner collection semi-angle are 25.0 mrad and 49.0 mrad, respectively.

**2.9 Resistivity.** The resistivity of $La_3Pd_2NaO_9$ was measured by the standard four-probe method. The crystals were too small to make four contacts, thus, we ground the crystals into powder and pressed them into a pellet. Then, the pellet was cut into two parts, one part for resistivity measurement and the other part was annealed in the high-pressure furnace at 420 °C and 70 bar $O_2$ for 3 days. The densities of the pellets before and after annealing were about 65% and 73% of single crystals, respectively. The pellet was cut into long and thin bars for making electrodes with DuPont silver paste. Resistivity was measured in the range of 160-300 K (four-probe method) and 100-300 K (two-probe method) on warming at a rate of 2 K/min on a Quantum Design magnetic property measurement system (MPMS3) with Electrical Transport Option without external magnetic field.

**2.10 Magnetic Susceptibility.** Magnetic susceptibility of the samples was measured using a Quantum Design MPMS3 VSM. Crystals of 32.8 mg were ground into powders and pressed into a pellet, and then glued onto the quartz rod using GE varnish. ZFC-W (zero-field cooling and data collected on warming), FC-C (field cooling and data collected on cooling), and FC-W (field cooling and data collected on warming) data were collected between 1.8 and 300 K under an external magnetic field of 1 T. The sample was cooled to 10 K at a rate of 10 K/min in zero field and held at 10 K for 10 min and then cooled to 1.8 K at a rate of 2 K/min and held at 1.8 K for 10 min. Then, we applied magnetic field to 1 T and magnetic susceptibility was recorded on warming (3 K/min). In the FC-C and FC-W procedures, the magnetic susceptibility was also recorded (3 K/min) in a fixed field of 1 T. Magnetization data as a function of the magnetic field were collected between −7 to +7 T with a sweep mode of 300 Oe/s at 1.8, 150, and 300 K. For comparison, the as-grown single crystals were also measured to explore if sodium in the material absorbs moisture that might result in different magnetic properties.

## 3. Results and Discussion

**3.1 High $pO_2$ Synthesis and Crystal Growth.** We have been exploring the possibility to synthesize R-P palladates with n > 1 at ambient pressure using the flux method. Initially, we tried the carbonate fluxes, hydroxide fluxes and chloride fluxes because R-P nickelates were synthesized using such fluxes;[56-61] however, we did not obtain any palladates with palladium valence of larger than +2 (see **Table S2** and **Figures S1-S5**). The failure to obtain high-valence Pd is probably due to insufficient oxidization; flux growth under high oxygen pressure may work. Among the above-mentioned fluxes, the melting points of carbonates are high (850 °C for $Na_2CO_3$ and 890 °C for $K_2CO_3$). In contrast, chlorides are corrosive at high oxygen pressure and it's very difficult to find a suitable crucible (Pt and $Al_2O_3$ did not work) and chamber. Molten alkali hydroxides have been utilized as fluxes for the growth of oxide crystals of the platinum group metals (Ru, Os, Ir, Rh, Pd, Pt) containing lanthanide metal.[62] It has been reported that divalent palladium oxides were synthesized using NaOH or KOH flux under normal pressure.[63-65] We expected to synthesize palladium oxides with higher oxidation states under high $pO_2$ conditions. Among hydroxide fluxes, we started with the NaOH-KOH (molar ratio 1:1) system because it has a lower melting point of ~170 °C, and single crystals of $NdNiO_3$ were successfully grown by our group.[66]

In our initial exploration, $La_2O_3$ and Pd were mixed according to the molar ratio 1:2, and then it was mixed with flux at a weight ratio of 1:20. The high-pressure furnace was kept at a temperature of 400, 410, or 420 °C under a $pO_2$ of 10, 30, 50, or 70 bar for 24 or 40 hours, and then cooled down to 300 °C at a rate of 10 °C/h (see specific growth conditions in **Table S1**). XRD analysis revealed the formation of perovskite phase at a $pO_2$ of ≥ 50 bar and a dwelling temperature of 420 °C (**Figure S6**). However, the positions of the XRD peaks are apparently shifted compared with the peaks of $LaPdO_3$ in the database (PDF #04-012-7471), indicating changes of lattice parameters. We then optimized the growth conditions, including weight ratio of raw materials to fluxes, dwelling time, and cooling rate in order to obtain pure phase. At a 1:10 weight ratio of raw materials to fluxes, dwelling temperature of 420 °C for 48 h, and a cooling rate of 3.5 °C/h, black powders with tiny but shiny crystals of ~5 μm were obtained. We also noticed several weak peaks at $2\theta$ angles smaller than 20° (**Figure S7**) that cannot be indexed using the perovskite structure of $LaPdO_3$. We performed LeBail fit (**Figure S8**) on the XRD of the crystals, and the results showed remarkable changes in the lattice parameters compared to $LaPdO_3$.[49] These results suggest that the as-grown single crystals are probably not $LaPdO_3$.

We continued to optimize the dwelling time and cooling rate in order to improve the dimensions and quality of crystals. The experimental details are shown in **Table S1**, and the XRD patterns are shown in **Figure S9**. No increase in the size of the crystals was found by extending the dwelling time from 48 h to 60 h. Thus, we used 48 h as the optimal dwelling time. We further slowed down the cooling rate to 2 °C/h, and finally obtained shiny crystals with dimensions of ~20 μm on edge (**Figure 1**). Unfortunately, the issue of crucible corrosion was much worse using slower cooling rates and long growth time, preventing us from growing larger single crystals.

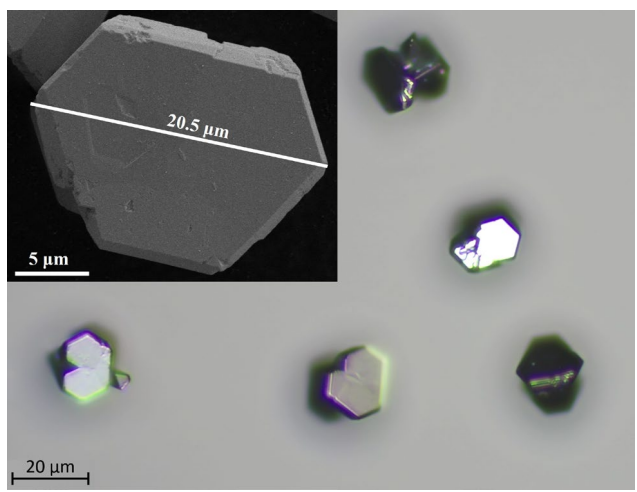

**Figure 1.** A typical photo of as-grown single crystals using a microscope. Inset shows a scanning electron microscope image.

**3.2 Determination of composition.** In order to determine the composition and find out the underlying mechanism of the big changes in the lattice parameters compared with perovskite $LaPdO_3$,[49] we performed EDS and ICP measurements. For EDS measurements, we selected all possible elements in our raw materials including La, Pd, O, Na, K and Al. We found no Al in the crystals although we used $Al_2O_3$ crucibles, and we found the presence of a significant amount of Na. **Table S3** lists the EDS data and **Figure S10** presents the raw data. To facilitate the analysis of the data, we normalized the La Atomic% to 1 and averaged the four sets of data. We found that the atomic ratio of La: (Pd+Na) is approximately 3: 3 and Pd: Na is 2: 1. Thus, the chemical formula of the crystal can be expressed as $La_3Pd_2NaO_x$, where ~1/3 of Na substitutes Pd compared to LaPdO$_3$. Considering that EDS only measures the surface with several μm in depth, we also measured the La, Pd and Na elements in the crystals by dissolving the crystals into acid using ICP-MS (see **Table S4**), which is an accurate method for determining the concentration of the elements in the bulk form. Five sets of samples containing three different concentrations were measured. Like the analysis of the EDS data, the molar concentration of La was normalized to 1 μmol/L and the five sets of data were averaged. The results are in excellent agreement with the EDS.

**3.3 XPS analysis.** We carried out XPS measurements to determine the valence of Pd. We first performed a full spectrum scan, and the results are shown in **Figure S11**. Clearly, no Al and K were found. Characteristic peaks from La, Pd, Na, and O are seen, which is in agreement with the EDS and ICP results. Scans near the C1s (280-290 eV) and the Pd3d (330-350 eV) binding energy regions were performed to obtain higher resolution spectrograms. Binding energy values were calibrated using the 284.8 eV peak of C1s. **Figure 2** shows the XPS spectrum of the Pd3d core level. As can be seen, the Pd3d spectrum shows a doublet Pd3d$_{5/2}$ and Pd3d$_{3/2}$ separated by 5.22-5.26 eV, in agreement with previous studies.[67] The two peaks of Pd3d$_{5/2}$ are located at 337.00 eV and 338.95 eV, respectively. The binding energies were reported to be about 336.1-337.1 eV for Pd$^{2+}$ and 337.7–339.3 eV for Pd$^{4+}$.[67-70] Thus, the peak at 337.00 eV indicates Pd$^{2+}$, while the peak at 338.95 eV corresponds to Pd$^{4+}$. The ratio of peak area of 3d$_{5/2}$ to 3d$_{3/2}$ for Pd$^{4+}$ is 1.63, and 1.52 for Pd$^{2+}$; both of them are consistent with the theoretical ratio of 3: 2. The ratio of peak area for Pd$^{4+}$ equal to 77%, indicating that the predominant oxidation state of Pd in the sample is +4, which aligns with our expectations. The presence of Pd$^{2+}$ suggests that our sample does not conform strictly to stoichiometry, possibly due to the existence of oxygen vacancies. Based on the amount of Pd$^{4+}$ (77%) and Pd$^{2+}$ (23%), and the atomic ratio of La/Pd/Na=3:2:1 from EDS and ICP, the oxygen content is estimated to be 8.55. Therefore, we estimate the chemical formula of the as-grown crystals to be La$_3$Pd$_2$NaO$_{8.55}$, which contains 5% oxygen deficiencies from stoichiometry (La$_3$Pd$_2$NaO$_9$).

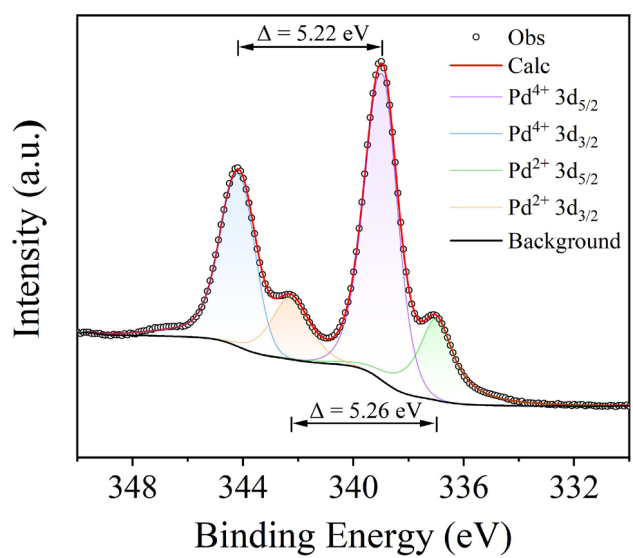

**Figure 2.** The XPS spectrum of the Pd-3d core level.

**3.4 Structural determination.** The X-ray powder diffraction pattern matches with a perovskite structure except several peaks at low angles, indicating that the crystal structure is perovskite or distorted perovskite. In principle, single-crystal X-ray diffraction is the best way to determine crystal structure; however, the as-grown crystals are too small, no good diffraction data were obtained via in-house X-ray single crystal diffraction. We thus performed Rietveld refinement on the in-house X-ray powder diffraction data by replacing 1/3 of Pd in the structural model of LaPdO$_3$ (space group *Pbnm*).[49] To ensure the accuracy of the refinement, we only refined a joint thermal parameter. The refinement (see **Figure S12**) converged to $R_{wp}$ = 9.86%, $R_{exp}$ = 5.18%, and GOF = 1.90 with lattice parameters of $a$ =

5.5729(2) Å, $b$ = 5.8086(2) Å and $c$ = 7.9657(2) Å. Apparently, several diffraction peaks at $2\theta < 20°$ are not accounted by this disordered model. There are two possibilities: (1) These peaks, consistently appeared under different experimental conditions (**Figures S7** and **S9**), are intrinsic to our as-grown single crystals. The single crystals are black and the morphologies observed using a microscope are pretty homogeneous, indicating probably a single phase. Along this line, the appearance of these weak peaks at low angles suggests a supercell compared to perovskite cell. (2) These weak peaks belong to impurities. We performed search-match using the latest powder diffraction database [ICDD-PDF-5+ (2025)], but found no match.

To determine the crystal structure, we carried out synchrotron X-ray single crystal diffraction at Beamline 13 at the Advanced Photon Source, Argonne National Laboratory. **Figure 3** shows the reconstructed ($h0l$) planes based on the perovskite unit cell of $a \sim 5.6$ Å, $b \sim 5.8$ Å and $c \sim 8.0$ Å. Notably, reflection peaks locating at $\boldsymbol{q}$ = (1/3, 0, 1/3) are clearly seen, indicating that the unit cell should be nine times large. In contrast, no superlattice diffraction peaks are observed on the ($0kl$) or ($hk0$) planes (**Figure S13**), possibly because they are too weak. The presence of the superlattice peaks suggests that Na atoms in the crystal likely form an ordered pattern with Pd. We then attempted to solve the crystal structure using the nine-times large supercell; however, we failed due to insufficient data.

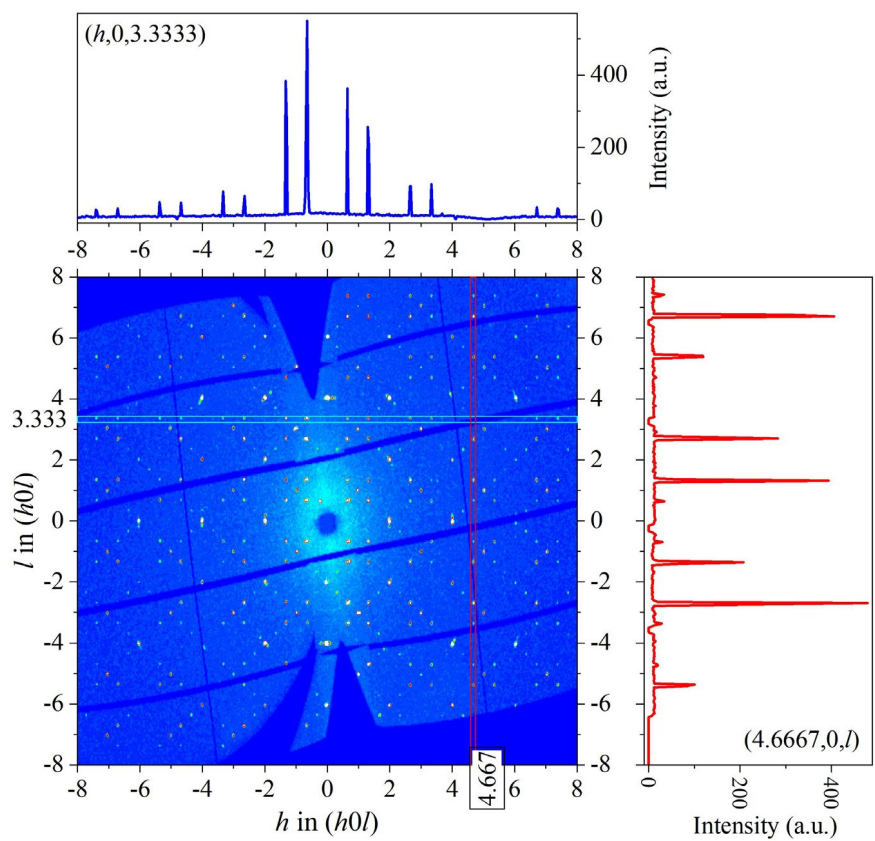

**Figure 3.** The reconstructed ($h0l$) planes based on the perovskite unit cell of $a \sim 5.6$ Å, $b \sim 5.8$ Å and $c \sim 8.0$ Å.

By lowering the symmetry to monoclinic $P2_1/c$, we succeeded in solving the crystal structure with lattice parameters of $a$ = 16.777(5) Å, $b$ = 5.8311(17) Å, $c$ = 9.759(3) Å, and $\beta$ = 124.969(3)°. Details of the crystal parameters and structure refinement are summarized in **Tables 1** and **2**. As shown in **Figure 4(a-c)**, the structure is a distorted perovskite. The asymmetric unit contains three La atoms, four Pd/Na positions and nine oxygen atoms (see **Table 2**). All Pd/Na atoms are coordinated by six oxygen atoms with bond length ranging from 1.981(6)-2.204(6) Å, forming octahedral environment, as shown in **Figure 4(d)**. These octahedra, where the Pd/Na atoms deviate from the center, share corners with each other. During the refinement process, we observed that some degree of Na disorder persists. However, Na preferentially occupies the Pd3 and Pd4 sites (see **Table 2**), demonstrating an overall charge-ordered state. We used the average Pd-O and La-O bond lengths, which are La-O = 2.589 Å, Pd(1)-O = 2.044 Å, Pd(2)-O = 2.041 Å, Pd(3)-O = 2.187 Å, Pd(4)-O = 2.183 Å, to estimate the tolerance factor. The Pd(1) and Pd(2) sites are primarily occupied by Pd, while the Pd(3) and Pd(4) sites are mainly occupied by Na. Therefore, the tolerance factor is calculated to be 0.90 based on the average bond lengths of Pd(1)-O and Pd(2)-O, and 0.84 based on the average bond lengths of Pd(3)-O and Pd(4)-O. The tolerance factor in LaPdO$_3$ is calculated to be 0.87. We note that the tolerance factor for the sites primarily occupied by Pd is slightly larger than that of LaPdO$_3$, while the tolerance factor for the sites primarily occupied by Na is smaller than that of LaPdO$_3$. However, all of them significantly deviate from 1, which is consistent with our results that the material is difficult to synthesize at ambient pressure and high oxygen pressure is needed. As shown in **Figure 4(e)**, within the (*h0l*) plane, along the *a*-axis direction, Pd and Na atoms essentially exhibit an ordered arrangement of two Pd layers followed by one Na layer. **Figure 4(f)** compares the unit cells of perovskite (black line rectangle), 3×3 supercell of the perovskite structure (dash rectangle) and our monoclinic *P*2$_1$/*c* model (blue parallelogram). We have attempted to topotactically reduce our as-grown sample using CaH$_2$, and we failed. Now this can be easily understood because of the incorporation of Na in the crystal structure. Along this line, we are exploring other fluxes to grow LaPdO$_3$ without Na substitution.

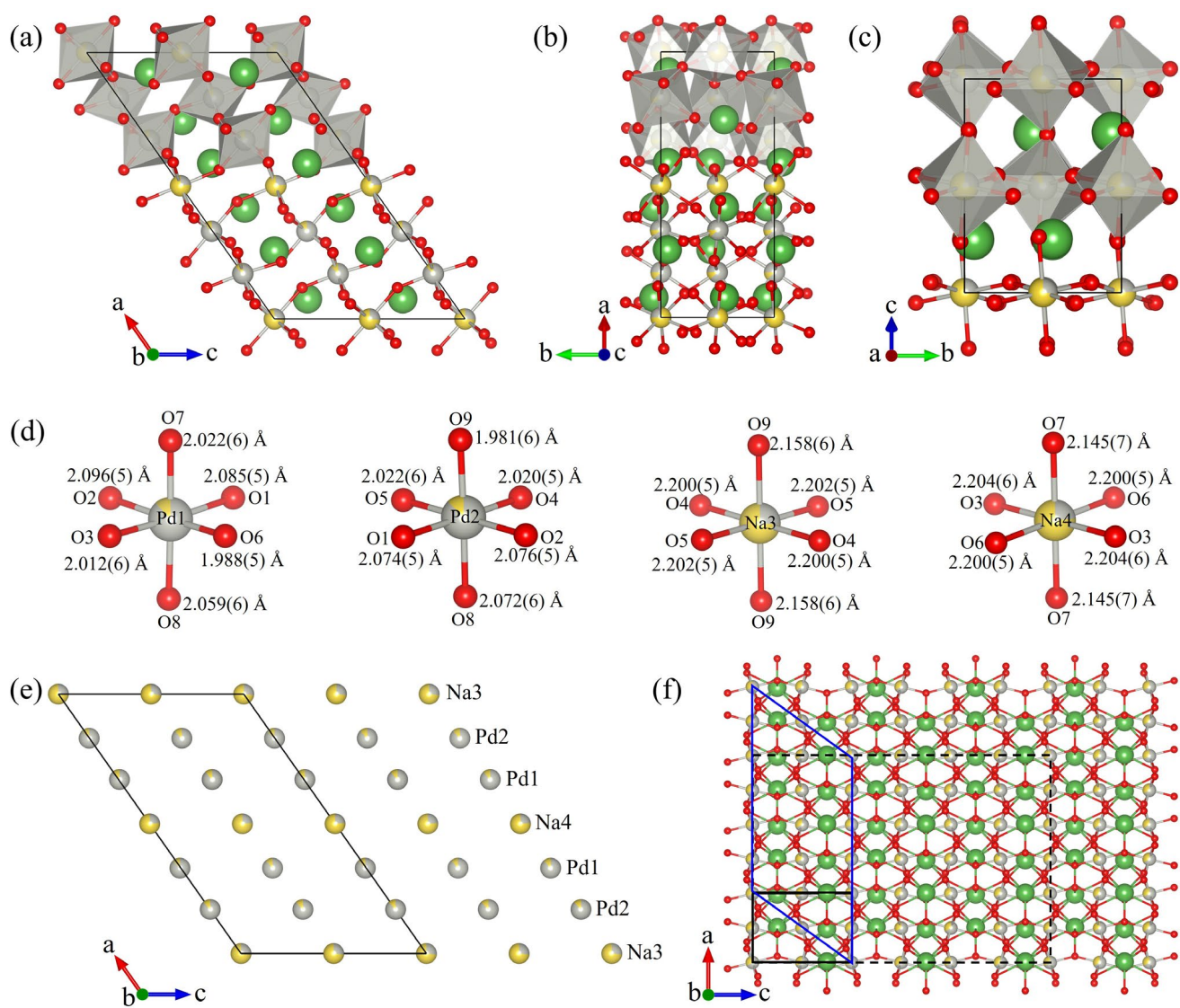

**Figure 4.** Crystal structure. The gray spheres represent palladium atoms, the yellow spheres represent sodium atoms, the red spheres represent oxygen atoms, the green spheres represent lanthanum atoms. (a-c) Monoclinic *P*2$_1$/*c* model obtained by SXRD. Note the model is shown in half polyhedral and half ball-and-stick style. (d) Pd/Na-O octahedral environments with bond lengths labeled of the monoclinic *P*2$_1$/*c* model. (e) Charge ordering of Na and Pd. Na and Pd

are disordered, and the dominant one is labeled. (f) The unit cells of perovskite structure of Na disordered occupancy obtained by Rietveld refinement on PXRD (black line rectangle), 3×3 supercell of the perovskite structure (dash rectangle) and monoclinic $P2_1/c$ model (blue parallelogram).

**Figure 5** shows the powder diffraction data (**experiment 13** shown in **Table S1**) as well as Rietveld refinement using the single crystal structural model obtained from SXRD (see **Figure S14** for zoom-in of low-angle data $2\theta$ = 12-25º and high-angle data $2\theta$ = 80-100º). The refinement converged to $R_{wp}$ = 8.52%, $R_{exp}$ = 5.18%, and GOF = 1.64 with lattice parameters of $a$ = 16.7184(4) Å, $b$ = 5.80857(13) Å, $c$ = 9.7216(11) Å, and $\beta$ = 124.975(10)°. It can be clearly seen that all peaks (particularly the low-angle peaks) were well fitted, confirming the single-crystal structure model. To explore whether the content of Na substitution for Pd would be different under different experimental conditions, we also performed Rietveld refinements on the powder diffraction data of the products from **experiments 4, 9, and 10** shown in **Table S1** and list the results in **Table S5**. Powder diffraction data and Rietveld refinement are shown in **Figure S15**. The lattice parameters obtained were basically the same (**Table S5**), indicating that the growth results are consistent and the material is a line compound.

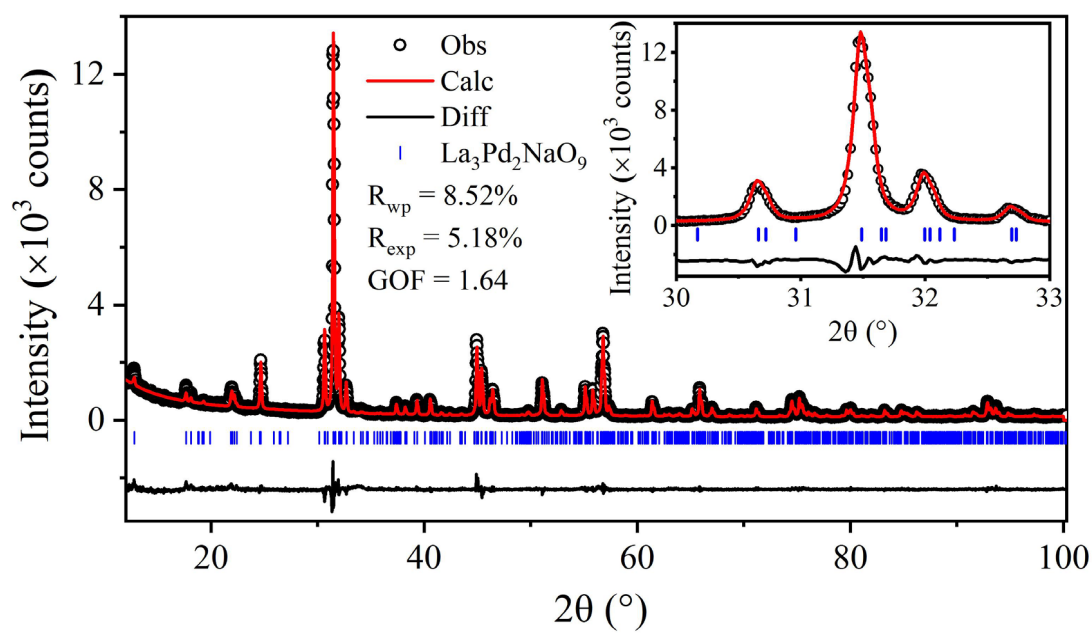

**Figure 5.** PXRD and Rietveld refinement using the single crystal structural model obtained by SXRD data. Obs: observed intensity; Calc: calculated intensity; Diff: difference.

**3.5 STEM analysis.** We investigated the local structure of $La_3Pd_2NaO_9$ single crystals using scanning transmission electron microscopy (STEM). Typical high-angle annular dark-field (HAADF)-STEM images along the [130] and [103] projections are shown in **Figure 6** (see **Figure S16 and S17** for wide range images). The $La_3Pd_2NaO_9$ single crystals exhibits a nice ordered alternating stacking of La-O and Pd/Na-O layers on the scale of tens of nanometers, which is consistent with the single crystal structure and indicates excellent sample homogeneity.

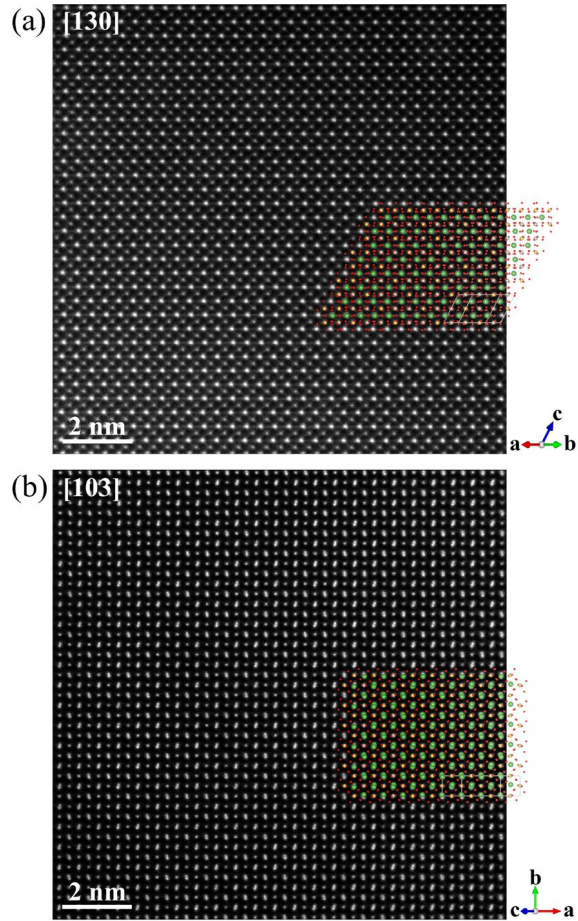

**Figure 6.** Typical atomic-scale HAADF-STEM images with overlaid single crystal structural model of $La_3Pd_2NaO_9$. (a) along the [130] projection. (b) along the [103] projection.

**3.6 Electrical resistivity.** Our as-grown crystals are too small to put four leads on, thus we pulverized the crystals to polycrystalline powders and pressed into pellets for electrical resistance measurements. **Figure 7** shows the resistivity as a function of temperature under zero magnetic field, we observed a semiconducting behavior. The as-grown sample shows variable range hopping (VRH) transport at low temperature expressed by $\rho \propto \exp[(\frac{T_0}{T})^{1/(d+1)}]$, where $d$ is the dimension of the system.[71] The good linear fit of the low temperature data at $d = 3$ indicates that the sample charge conducts in three dimensions, consistent with the type of transport in the three-dimensional structure.

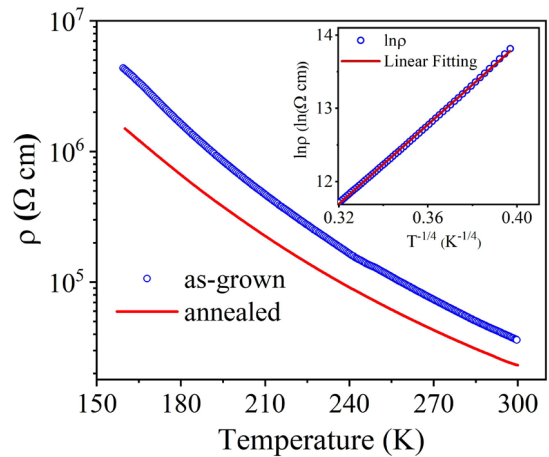

**Figure 7.** Electrical resistivity of polycrystals. Inset figure shows the $\ln\rho$ of as-grown sample plotted as function of $T^{-1/4}$.

In perovskite, oxygen content has a significant impact on transport properties.[72, 73] To increase oxygen content, we annealed our pellets at 420 °C and 70 bar of oxygen for three days. **Figure 7** shows the resistivity of annealed sample (see **Figure S18** for wide range of resistivity measured by the two-probe method). It is still semiconducting, although the resistivity becomes smaller compared with as-grown samples. The resistivity of our sample is seven orders of magnitude larger compared to the LaPdO$_3$ tested by Penny G.[53] The insulating behavior of La$_3$Pd$_2$NaO$_9$ can be understood due to the formation of a charge-ordered pattern between Na and Pd ions. Similar phenomenon has been reported in the trilayer square planar system La$_4$Ni$_3$O$_8$.[12]

**3.7 Magnetic susceptibility and magnetization. Figure 8(a)** shows the magnetic susceptibility of polycrystals as a function of temperature ranging from 1.8 to 300 K. The sample exhibits paramagnetic behavior. The high-temperature magnetic susceptibility is analyzed. A Curie-Weiss fit to the ZFC-W data in the range of 200-300 K using $\chi = \chi_0 + C/(T- \theta)$, where $\chi_0$, C and $\theta$ are the temperature independent contribution, Curie and Weiss constants, respectively, yields $\chi_0$ = 1.74×10$^{-5}$ emu mol$^{-1}$ Oe$^{-1}$, C = 0.035 emu K mol$^{-1}$ Oe$^{-1}$ and $\theta$ = -269 K. The resulting effective moment is 0.53 $\mu_B$ per Pd ion.[74] The negative value of the Weiss constant indicates antiferromagnetic interactions. According to the XPS results, the oxidation state of Pd in La$_3$Pd$_2$NaO$_9$ is mostly +4 with a small amount of +2. In the octahedral field, the $d$ orbital splits into three $t_{2g}$ orbitals and two $e_g$ orbitals, and Pd$^{4+}$ likely possess the $t_{2g}^6$ electronic configuration, thus it is expected to be diamagnetic. Meanwhile, Pd$^{2+}$ has a d$^8$ electron configuration, similar to Ni$^{2+}$, and may exist in two spin states: low-spin ($S$ = 0) and high-spin ($S$ = 1).[5] Therefore, the observed paramagnetism may be attributed to the presence of a small fraction of high-spin Pd$^{2+}$ caused by oxygen defects in the sample. Similar phenomena have been reported in K$_2$PdO$_3$,[75] Zn$_2$PdO$_4$,[76] KPd$_2$O$_3$,[77] and BiPd$_2$O$_4$,[78] and in these cases defects were proposed to be the reason for paramagnetism.

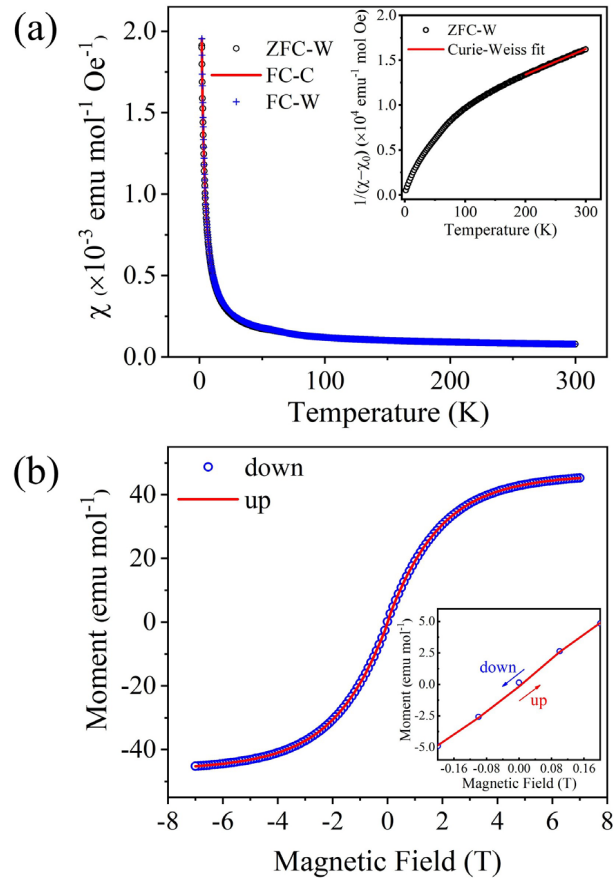

**Figure 8.** (a) Magnetic susceptibility of La$_3$Pd$_2$NaO$_9$. Inset figure shows the Curie-Weiss fit of ZFC-W. (b) Magnetization at 1.8 K as a function of magnetic fields (blue: +7 to −7 T, red: −7 to +7 T).

**Figure 8(b)** shows the magnetization of polycrystals at 1.8 K as a function of magnetic fields ranging from 7 T to -7 T. Magnetization as a function of magnetic fields ranging from 7 T to -7 T at 150 K and 300 K are shown in **Figure S19**. At low field, paramagnetic behavior is observed. In contrast, diamagnetic signal is clearly seen at high magnetic field, consistent with the expected $S = 0$ ($t_{2g}^6$ configuration) of $Pd^{4+}$. Considering that sodium exists in our samples and ground powders might absorb moisture, thus affecting the magnetic properties, we also measured the as-grown single crystals. The magnetic results were consistent with those of the powders, and the difference in fitting parameters are relatively small (**Figure S20**).

## 4. Conclusions

We have successfully grown single crystals of a high-valent palladate $La_3Pd_2NaO_9$ with dimensions of up to 20 μm on the edge at 420 °C and 70 bar oxygen pressure using the NaOH-KOH (molar ratio 1:1) flux system. The as-grown crystals exhibit shiny natural facets determined by its crystal symmetry. EDS, ICP and XPS measurements revealed that the chemical formula of the as-grown single crystals is $La_3Pd_2NaO_{9-\delta}$ ($\delta$ = 0.45). Synchrotron X-ray single-crystal diffraction data reveals that the crystal belongs to the monoclinic $P2_1/c$ with lattice parameters of $a$ = 16.777(5) Å, $b$ = 5.8311(17) Å, $c$ = 9.759(3) Å, and $\beta$ = 124.969(3)°. STEM measurements demonstrate excellent sample homogeneity. Resistivity measurements of $La_3Pd_2NaO_9$ indicate that it is an insulator, which can be attributed to charge ordering induced by the ordered Na and Pd. The magnetic susceptibility and magnetization data show a paramagnetic behavior, which is inconsistent with the expected $t_{2g}^6$ configuration ($S = 0$) for $Pd^{4+}$. The anomalous magnetic properties may be due to the presence of a small fraction of high-spin $Pd^{2+}$ caused by oxygen defects in the sample. The successful preparation of $La_3Pd_2NaO_9$ single crystals with a high-valent Pd oxidation state at a relatively low temperature and oxygen pressure provides us with a new approach to explore the preparation of new R-P palladates.

## Associated content

**Supporting Information.** The Supporting Information is available free of charge at XXX.

Tables of synthesis and growth conditions of single crystals, EDS, ICP, Rietveld refinement; and figures of powder diffraction data, LeBail fit, EDS raw data, XPS full spectrum, Rietveld refinements using various models, the reconstructed (*0kl*) and (*hk0*) planes, HAADF-STEM, electrical resistivity, and magnetic properties of as-grown crystals (PDF)

## Accession Codes

CCDC 2445228 contains the supplementary crystallographic data for this paper. These data can be obtained free of charge via www.ccdc.cam.ac.uk/data_request/cif, or by emailing data_request@ccdc.cam.ac.uk, or by contacting The Cambridge Crystallographic Data Centre, 12 Union Road, Cambridge CB2 1EZ, UK; fac: +44 1223 336033

## Author Information

Corresponding Author

**Junjie Zhang** – Institute of Crystal Materials, State Key Laboratory of Crystal Materials, Shandong University, Jinan 250100, Shandong, China; orcid.org/0000-0002-5561-1330; Email: junjie@sdu.edu.cn

Authors

**Qingqing Yang** – Institute of Crystal Materials, State Key Laboratory of Crystal Materials, Shandong University, Jinan 250100, Shandong, China

**Ning Guo** – CAS Key Laboratory of Standardization and Measurement for Nanotechnology, CAS Center for Excellence in Nanoscience, National Center for Nanoscience and Technology, Beijing 100190, China

**Tieyan Chang** – NSF's ChemMatCARS, The University of Chicago, Argonne, IL 60439, USA;

orcid.org/0000-0002-7434-3714

**Zheng Guo** – Institute of Crystal Materials, State Key Laboratory of Crystal Materials, Shandong University, Jinan 250100, Shandong, China

**Xiaoli Wang** – Institute of Crystal Materials, State Key Laboratory of Crystal Materials, Shandong University, Jinan 250100, Shandong, China

**Chuanyan Fan** – Institute of Crystal Materials, State Key Laboratory of Crystal Materials, Shandong University, Jinan 250100, Shandong, China

**Chao Liu** – Institute of Crystal Materials, State Key Laboratory of Crystal Materials, Shandong University, Jinan 250100, Shandong, China

**Lu Han** – Institute of Crystal Materials, State Key Laboratory of Crystal Materials, Shandong University, Jinan 250100, Shandong, China

**Feiyu Li** – Institute of Crystal Materials, State Key Laboratory of Crystal Materials, Shandong University, Jinan 250100, Shandong, China

**Tao He** – Institute of Crystal Materials, State Key Laboratory of Crystal Materials, Shandong University, Jinan 250100, Shandong, China

**Qiang Zheng** – CAS Key Laboratory of Standardization and Measurement for Nanotechnology, CAS Center for Excellence in Nanoscience, National Center for Nanoscience and Technology, Beijing 100190, China

**Yu-Sheng Chen** – NSF's ChemMatCARS, The University of Chicago, Argonne, IL 60439, USA

**Author Contributions**
The manuscript was written through the contributions of all authors.

**Notes**
The authors declare no competing financial interest.

**Acknowledgments**
Work at Shandong University was supported by the National Natural Science Foundation of China (Grant Nos. 12374457 and 12074219). J. Z. thanks Prof. Xutang Tao from Shandong University for providing valuable support and fruitful discussions. J. Z. and Q. Y. would like to acknowledge the technical support from Shandong University Core Facilities Sharing Platform. J. Z. and Q. Y. thank Menghui Chu for his help with Inductively Coupled Plasma Mass Spectroscopy, and Jiahui Lu for her help with Scanning Electron Microscopy and Energy Dispersive Spectroscopy.

For Table of Contents Only

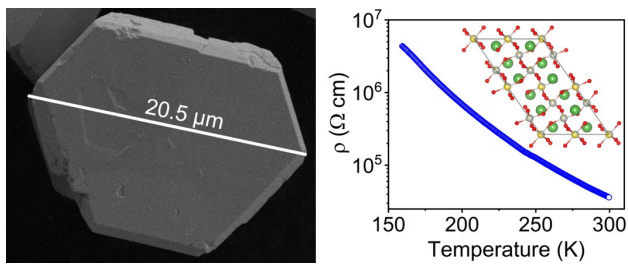